\documentclass[onecolumn,superscriptaddress,english,pre,showpacs,longbibliography]{revtex4-2}

\usepackage{amsmath}
\usepackage{graphicx}
\usepackage{dcolumn}
\usepackage{bm}
\usepackage{hyperref}
\usepackage{algorithm}
\usepackage{algorithmic}
\usepackage{color}
\usepackage{amssymb}
\usepackage{graphicx}
\usepackage{color}
\usepackage{mathrsfs}
\usepackage{float}
\usepackage{indentfirst}
\usepackage{txfonts}
\usepackage{booktabs} 
\usepackage{multirow}

\DeclareMathOperator\arctanh{arctanh}

\newcommand{\sumup}{\sum\limits}

\newcommand{\la}{\langle}
\newcommand{\ra}{\rangle}

\newcommand{\beq}{\begin{equation}}
\newcommand{\eeq}{\end{equation}}

\newcommand{\h}{h}
\begin{document}


\title{Solving Statistical Mechanics on Sparse Graphs with \\Feedback Set Variational Autoregressive Networks}


\author{Feng Pan}
\affiliation{
 CAS Key Laboratory for Theoretical Physics, Institute of Theoretical Physics, Chinese Academy of Sciences, Beijing 100190, China
}
\affiliation{
 School of Physical Sciences, University of Chinese Academy of Sciences, Beijing 100049, China
}
\author{Pengfei Zhou}
\affiliation{
 CAS Key Laboratory for Theoretical Physics, Institute of Theoretical Physics, Chinese Academy of Sciences, Beijing 100190, China
}
\affiliation{
 School of Physical Sciences, University of Chinese Academy of Sciences, Beijing 100049, China
}
\author{Hai-Jun Zhou}
\email{zhouhj@itp.ac.cn}
\affiliation{
 CAS Key Laboratory for Theoretical Physics, Institute of Theoretical Physics, Chinese Academy of Sciences, Beijing 100190, China
}
\affiliation{
 School of Physical Sciences, University of Chinese Academy of Sciences, Beijing 100049, China
}
\author{Pan Zhang}
\email{panzhang@itp.ac.cn}
\affiliation{
 CAS Key Laboratory for Theoretical Physics, Institute of Theoretical Physics, Chinese Academy of Sciences, Beijing 100190, China
}

\date{\today}

\begin{abstract}
We propose a method for solving statistical mechanics problems defined on sparse graphs. It extracts a small \textit{Feedback Vertex Set} (FVS) from the sparse graph, converting the sparse system to a much smaller
system with many-body and dense interactions with an effective energy on every configuration of the FVS, then learns a variational distribution parameterized using neural networks to approximate the original Boltzmann distribution. The method is able to estimate free energy, compute observables, and generate unbiased samples via direct sampling without auto-correlation. Extensive experiments show that our approach is more accurate than existing approaches for sparse spin glasses. On random graphs and real-world networks, our approach significantly outperforms the standard methods for sparse systems such as the belief-propagation algorithm; on structured sparse systems such as two-dimensional lattices our approach is significantly faster and more accurate than recently proposed variational autoregressive networks using convolution neural networks. 
\end{abstract}

\maketitle


\section{\label{sec:intro}introduction}
The underlying graph of interacting systems are often quite sparse, these include physics models defined on two-dimensional or three-dimensional lattices which encode physical interactions; social systems on networks which represent social relations~\cite{newman2018networks}; mathematical and information systems on sparse factor graphs which induce constraints over variables as in the satisfiability problems~\cite{Mezard2002} and in the low density error correcting codes~\cite{fossorier1999reduced,Mezard2009}.
The \textit{statistical mechanics} problems defined on such systems are of great importance: it can e.g. characterize phases and phase transitions in finite-dimensional physics model, describe dynamics of human behaviors, formulate posterior distribution in Bayesian inference, compute number of solutions of constraint satisfaction problems, and conduct message-passing algorithms for reconstructing signals in error-correcting codes, etc.
Without loss of generality, in this work we consider the prototype problem of statistical mechanics with $n$ spins sitting on a sparse graph $\mathcal {G}$. The joint probability distribution of a configuration $\mathbf s\in\{+1,-1\}^n$ follows the Boltzmann distribution
\begin{equation}
    p(\mathbf s)=\frac{1}{Z}e^{-\beta E(\mathbf s)},
\end{equation}
where $Z$ denotes the partition function, $\beta$ denotes the inverse temperature, and $E$ is the energy function. It is well known that computing the partition function, or equivalently the free energy 
\begin{equation}\label{eq:F}
    F = -\frac{1}{\beta}\ln Z = -\frac{1}{\beta}\ln\sum_{\mathbf s}e^{-\beta E(\mathbf s)},
\end{equation}
belong to the class of $\#$P complete problems, so it is hopeless to find polynomial algorithms to solve it in general. In statistical physics, one usually applies approximate (mean-field) methods for estimating an upper bounds of the free energy, or subjects to sampling methods which generate samples and calculate observables based on the samples.

The sparse systems (e.g. on social networks) we are interested in are usually large, the Markov-Chain Monte Carlo (MCMC) based sampling methods are consider to be not scalable. For systems with \textit{random sparsity}, such as on random graphs and real-world large networks, variational approximations based on the Bethe approximation~\cite{Bethe1935}, i.e., belief propagation (BP)~\cite{Yedidia2001}, have been widely applied. The key idea in the Bethe approximation, the conditional independence assumption, is exact when the system is a tree, and is usually a good approximation in random sparse systems. However, the approximation does not perform well when the system contains many short loops. Approximations such as Kikuchi loop expansions~\cite{kikuchi1951theory} have been proposed to partly take into account the effects of short loops, however it is still far from being optimal when the system has loops with different lengths such as lattices, which we call \textit{structured sparse} systems. In this kind of systems, tensor network based renormalization and block decimation approaches are very powerful, but unfortunately they work poorly if system does not contain many intrinsic low-rank (or low-entanglement) structures, and they do not apply when system contains long-range interactions.

Recently a variational method using neural networks, the Variational Autoregressive Network (VAN), is proposed for general statistical mechanics problems~\cite{wu2019solving}. On small dense systems, VAN uses multi-layered neural networks to describe a variational distribution, and achieves much better performance in estimating free energy, obtaining observables, and in sampling, when compared with traditional variational mean-field methods. However the VAN contains a huge number of parameters to optimize hence is not designed for large sparse systems at the first place. On structured sparse systems such as two-dimensional (2D) lattices, the issue can be eased by adopting convolution neural networks~\cite{DLbook} as a good structure prior. But on random sparse systems it is still not clear how to propose efficient convolution-like neural network structures for variational computations.


In this work extend the VAN approach to statistical mechanics problems on sparse graphs. Our approach converts a large sparse system to an equivalent but much smaller system composed of the vertices in a \textit{Feedback Vertex Set} (FVS) of the graph~\cite{zhou2013spin,zdeborova2016fast}. After the conversion, the reduced system is densely connected with many-body interactions with an effective energy on each configuration of FVS vertices. We then apply the VAN method to solve the statistical mechanics problem on the reduced system.

At a first glance, we converts a sparse system to a strongly correlated dense system which seems even harder than the original one. Indeed it is, in the sense that the converted system loses all the internal sparse structures of the original system. But notice that the key advantage of this converting is that the recently proposed neural-network based variational methods are particularly suitable for treating a small non-structured dense system. We will show in what follows that the VAN method can utilize recently developed machine learning techniques such as reinforcement learning, as well as advanced computational techniques such as GPUs, to target the converted problem on the compressed dense graph, and significantly outperforms existing methods operating on the original sparse system.

The rest of the paper is organized as follows. In Sec.\ref{sec:fvs} we introduce the feedback set of sparse graphs and how to convert the statistical mechanics problems on a sparse graph to the problem defined on its feedback set. In Sec.\ref{sec:fvsvan} we explain how to solve the converted statistical mechanics problem using variational autoregressive networks. In Sec.\ref{sec:ne} we conduct extensive experiments on spin glasses and inference problems to evaluate our approach against mean-field methods, VAN with convolution neural networks, and the recently proposed Expectation Propagation based Density Consistency method. We conclude in Sec.\ref{sec:con}.
\section{Feedback Set Variational Autoregressive Networks}
Our method is composed of two steps, the first step is converting the statistical mechanics problem in a sparse graph to a problem with many body interactions in a much smaller dense graph, using an approach based on the feedback set of the sparse graph which assign each configuration of the feedback set an effective energy; the second step of our method is applying the recently proposed autoregressive variational network using effective energies obtained in the first step to solve the statistical mechanics on the dense graph. In the following text we will introduce the two steps in detail.
\subsection{Compressing the sparsity using the feedback vertex set\label{sec:fvs}}

A feedback vertex set $\Omega$ of a graph is a set of vertices intersecting with every loop of the graph. With all vertices of the FVS removed, the graph becomes loop-free, hence is composed of trees~\cite{zhou2013spin}. We denote the forest after removing all the vertices of the FVS as $\mathcal{T}$. 
The free energy of the whole system~Eq.\eqref{eq:F} can be re-written in the following form by tracing over FVS variables and non-FVS variables separately:
\begin{equation}
\label{eq:Z}
F =  -\frac{1}{\beta}\ln\sum_{\mathbf s\in\Omega}\sumup_{\mathbf t\in \mathcal{T}}e^{-\beta E(\bm{s}, \bm{t})} 
= -\frac{1}{\beta}\ln\sum_{\mathbf s\in \Omega} e^{-\beta \widetilde{E}(\bm{s})} \; ,
\end{equation}
where $\mathbf s\in\{+1,-1 \}^{|\Omega|}$ denotes an assignment of variables on FVS, and $\mathbf t\in\{+1,-1 \}^{|\mathcal {T}|}=\{+1,-1 \}^{n-|\Omega|}$ denotes an assignment of variables in the forest out of the FVS.
$E(\bm{s}, \bm{t})$ denotes the energy of the original sparse system, and $\widetilde{E}(\bm{s})$ is the effective energy of a configuration $\bm{s}$ of the FVS vertices.  The effective energy can be written as
%
\begin{equation}
\label{eq:sumup}
\widetilde{E}(\bm{s}) 
 = E_{\Omega}(\bm{s}) + E_{\mathcal{T}}(\bm{s}) \; .
\end{equation}
%
Here $E_{\Omega}(\bm{s})$ denotes the sum of interaction energies (of the original sparse system) involving only vertices of the FVS, and 
\begin{equation}
\label{eq:tree energy}
    E_{\mathcal{T}}(\bm{s}) =  -  \frac{1}{\beta} \ln \bigg( \sum_{\bm{t}\in\mathcal{T}} e^{-\beta E(\bm{s}, \bm{t})} \bigg) \; ,
\end{equation}
is the (free) energy of the forest $\mathcal{T}$ given the configuration $\bm{s}$ of the FVS.

Observe that, once a configuration $\mathbf s$ of the FVS is determined, what left in the graph is a forest with boundary conditions fixed by $\mathbf s$. As a consequence, the summation over all variables in the forest as shown in Eq.~\eqref{eq:tree energy} can be calculated in linear time straightforwardly, i.e., by taking a sequence from leaf vertices (a leaf vertex of the graph is a vertex which has only one attached edge) of the forest to the root vertices. A detailed example about calculating free energy of the Ising model based on this method can be found in the appendix.

There are many possible FVS one can find for the graph $\mathcal G$. Finding the smallest one, which is called the minimum feedback vertex set problem, belongs to the class of {\it{NP-hard}} problems. In this work we employ an heuristic algorithm called CoreHD, proposed in ~\cite{zdeborova2016fast}, which is a fast and simple  heuristic method with $O(n)$ computation complexity.  The sizes of the FVS for the graph instances studied in this work are about $20\%$-$33\%$ of the original graph sizes. We note that it is possible to further reduce the FVS sizes through a more advanced message-passing method~\cite{zhou2013spin}.

Equation~\eqref{eq:Z} converts the partition-function calculation of a sparse system with size $n$ to that of a system of much smaller size $|\Omega|$. The conversion comes with price: the effective energy of the FVS indicates that there are complicated many-body interactions, thus good approximations on sparse graphs (e.g.,  Bethe approximation) fail completely in the FVS system. Indeed it is really difficult to find good approximations or variational ansatz for this dense and many-body interacting system in general, so we adopt a neural network based method.

\subsection{Variational autoregressive networks for statistical mechanics on the FVS\label{sec:fvsvan}}
The Variational Autoregressive Network~\cite{wu2019solving} has been proposed recently for statistical mechanics problems in general, and it is particularly powerful for densely connected structureless problems. 
Analogous to the mean-field methods, the VAN maintains a variational distribution $q_\theta(\mathbf s)$, and gradually updates parameters $\theta$ of the distribution to minimize the variational free energy. What is different from the canonical mean-field methods is that in the VAN, the variational distribution is represented using product of conditional probabilities
\begin{equation}\label{eq:autoregressive}
q_\theta(\mathbf s) = \prod_{i = 1}^n 
q_\theta(s_i \mid s_1, \ldots, s_{i - 1}),
\end{equation}
with all the conditional distributions parameterized by a neural network. 
The advantage of the autoregressive distribution as well as the neural network representation is the representation power. Theoretically it is able to represent any Boltzmann distribution given enough parameters in the hidden layers of the neural network, known as the \textit{representation theorem}~\cite{csaji2001approximation}, so is much more representative compared with mean-field ansats such as Bethe and Thouless-Anderson-Palmer~\cite{thouless1977solution}. However the price that VAN pays for the representation power is that the variational free energy is no longer analytically computable, as opposed to mean-field methods where the variational free energies and their derivatives over parameters can always be written out analytically. As a solution, the VAN generates many unbiased samples from the variational distribution with exact probability (known as direct sampling in ~\cite{bishop2006pattern}]), which is easily computable using Eq.~\eqref{eq:autoregressive}, then estimates the variational free energy using

\begin{equation}
    F_q = \sumup_{\bm{s}} q_{\theta}(\bm{s}) \Bigl[ \widetilde E(\bm{s})+ \frac{1}{\beta}\ln q_{\theta}(\bm{s}). \Bigr]\label{eq:Fq}
\end{equation}
Once the variational free energy is computed, in VAN we compute its gradients with respect to parameters $\theta$ using
\begin{equation}
\nabla_{\theta} F_q = 
\sum\limits_{\bm{s}} q_\theta(\bm{s}) 
\biggl\{ \Bigl[ \widetilde E(\bm{s}) + \frac{1}{\beta} \ln q_{\theta}(\bm{s})\Bigr]  \nabla_{\theta} \ln q_{\theta}(\bm{s}) \biggr\} \; ,
 \label{eq:gradient}
\end{equation}
which is known as \textit{reinforce} algorithm, in the machine learning literature~\cite{williams1992simple}. A short introduction to VAN and the learning algorithm can be found in the appendix (see also Ref.~\cite{wu2019solving} for more details). 
Using the gradients, we update parameters of the VAN gradually until the variational free energy stops decreasing. After that, the learnt VAN can be used to generate a massive number of configurations unbiasely and parallelly using the direct sampling, and then observables such as magnetizations and correlations can be computed straightforwardly using the configurations.


Our refer to the algorithm as FVSVAN, it takes the sparse graph $\mathcal G$ and energy function $E$ as input, and outputs the variational free energy, unbiased samples after the learning is finished. Given $\mathcal G$ and $E$, the FVSVAN is sketched as follows: 
\begin{enumerate}
    \item {Extract a FVS $\Omega$ from the graph $\mathcal G$ using the CoreHD algorithm~\cite{zdeborova2016fast}.}
    \item {Construct VAN to express the variational distribution $q_\theta(\mathbf s)$ for $s\in\Omega$, initialize neural networks in VAN randomly.}
    \item{Do direct sampling to obtain unbiased samples $\{\mathbf s\}$.}
    \item{Compute effective energy $\widetilde E(\mathbf s)$, probability $q(\mathbf s)$ for each sample $\mathbf s$ using Eq.~\eqref{eq:sumup} and Eq.~\eqref{eq:autoregressive} respectively, then compute variational free energy $F_q$ according to Eq.~\eqref{eq:Fq}. }
    \item {Compute gradients of the variational free energy with respect to parameters according to Eq.~\eqref{eq:gradient}, then update the parameters (i.e. weights of neural network) of the variational distribution using the gradients.}
    \item {Go to step 3. if the variational free energy has not converged.}
\end{enumerate}

\section{\label{sec:ne}Numerical experiments}

To evaluate the ability of our approach in estimating the free energy and in computing physical quantities in sparse systems, and to evaluate our method against existing approaches, we perform extensive experiments on sparse graphs using spin glasses and inference problems.

There are two kinds of sparsities we usually encounter in statistical mechanics problems: random sparsity in e.g random graphs and real-world sparse networks, and structured sparsity in e.g. $2$D and $3$D lattices. For random graphs and large real-world networks, the Bethe approximation based methods such as the belief propagation are considered to be standards approaches. A recently proposed approach, Density Consistency, (DC)~\cite{braunstein2019loop} which is based on the Expectation Propagation~\cite{minka2013expectation} algorithm, has been shown to outperform BP in calculating physical quantities, so we also take this approach as a baseline for comparison in our experiments. On the lattice systems, recently proposed convolution based VAN~\cite{wu2019solving} takes the structure prior of $2D$ graphs by adapting the convolution neural networks, and largely outperforms mean-field approximations so we mainly focus on the comparison to the VAN with convolution neural networks operating on the whole lattice. 
In our experiments we consider two kinds of problems. The first one is a representative model of spin glasses where the tasks include estimating the tree energy, and computing the correlation functions; the second problem we consider is an inference problem on sparse graphs known as \textit{censored stochastic block model} ~\cite{saade2015spectral}, where the task is to solve the posterior distribution (which is the Boltzmann distribution) in the Bayesian inference and infer the planted partition using observations on each edge of the graph.

\subsection{Spin glasses on random graphs}

\begin{figure}[!t]
\centering
\includegraphics[width=0.6\columnwidth]{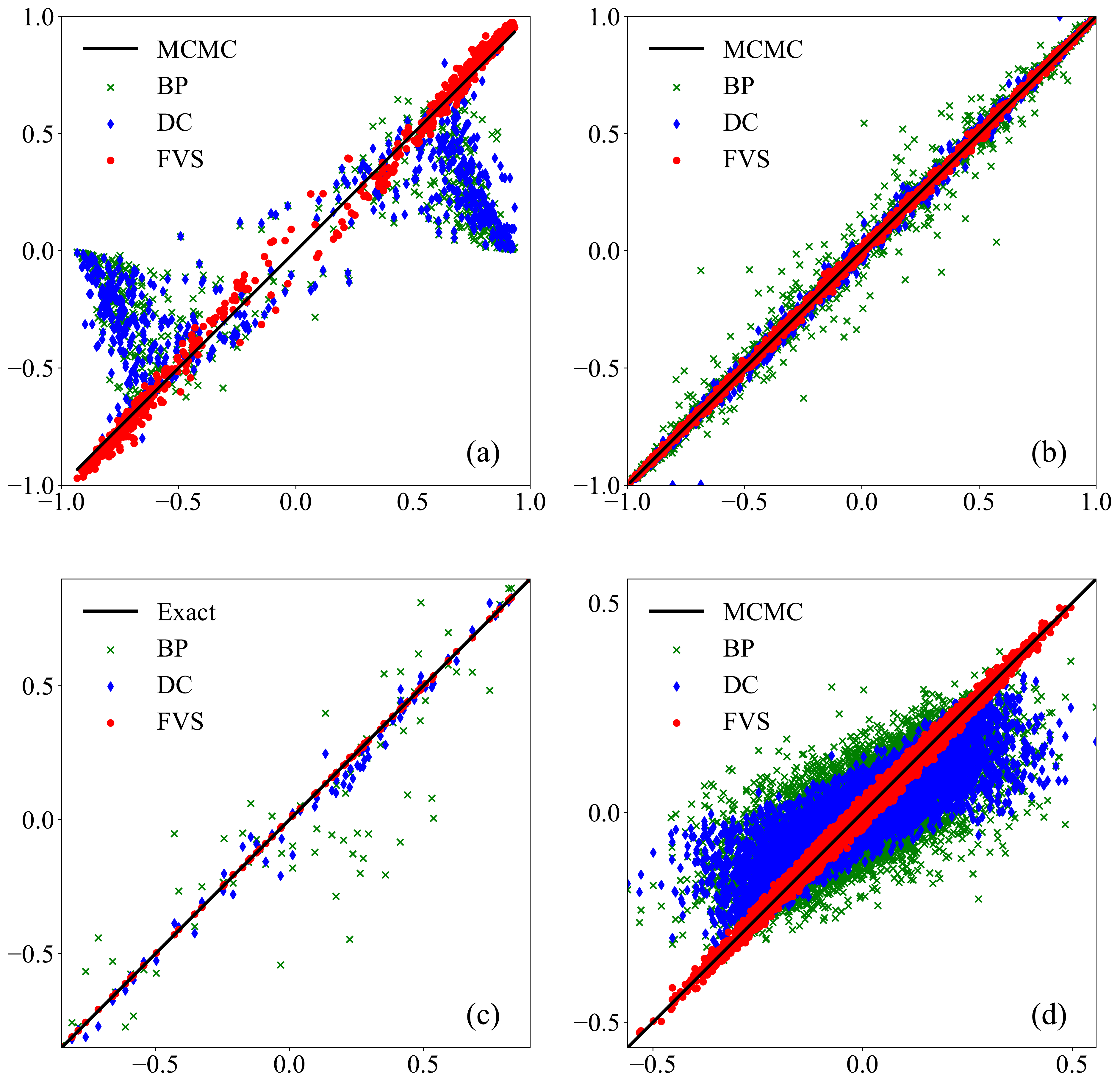}
\caption{\label{fig:correlation} Correlations of spin glass models obtained by our method (FVS), belief propagation (BP) and density consistency (DC) compared with data given by MCMC running for a long time ($5\times 10^5 n$ steps) on various of graphs. (a) Viena-Bray spin glass model~\cite{Viana1985} on random regular graph, $n=300$ spins, degree $4$, $\beta=0.8$, couplings $J_{ij}\in\{+1,-1\}$ with $P(J_{ij}=1)=P(J_{ij}=-1)=\frac{1}{2}$; (b) Erd\"os-R\'enyi random graphs, $n=1100$ spins, average degree  $3$, $\beta=0.8$ and couplings are Gaussian random variables with zero mean and unit variance; (c) The model is the same as that of (b) but on the real-world karate club network~\cite{zachary1977information}, with $n=34$ variables, average degree $2.29$ and $\beta=0.54$; (d) The same as (c) but on the real-world political blogs network~\cite{adamic2005political}, with $n=1490$ variables, average degree $11.21$ and $\beta=0.1$.}
\end{figure}

We first consider the Ising $\pm J$ spin glasses on random graphs, the so-called Viena-Bray spin glass model~\cite{Viana1985}, with distribution of couplings following $P(J_{ij}=1)=P(J_{ij}=-1) = 1/2.$
We evaluate performance of our approach by comparing correlations obtained by our method to those obtained using MCMC for a very long time ($5\times 10^5 n$ steps, and we observed that longer time simulation gives almost identical results), which is considered to be very accurate. We also compare our results against correlations obtained by the BP and DC algorithms. 
The results are shown in Fig.~\ref{fig:correlation}.

In Fig.~\ref{fig:correlation}(a) the graph is regular random graph with degree of every node being $4$. We set $\beta=0.8$ which is beyond the spin glass transition and the system is at the spin glass phase. We can see from the figure that the correlations given by our method are very close to the MCMC results, while the BP and DC results deviate significantly from the MCMC data. 
In Fig.~\ref{fig:correlation}(b) the spin glass model has couplings $J$ following Gaussian distribution with unit variance, and on an Erd\"os-R\'enyi random graph. We can see that the results of BP and DC of are much better than those in Fig.~\ref{fig:correlation}(a), 
but they are still much less accurate than our method. 
In Fig.~\ref{fig:correlation}(c) and (d) we show results of our algorithm on two classic real-world networks, the Karate club network~\cite{zachary1977information} and the network of political blogs~\cite{adamic2005political}. On the Karate club we see that our method is almost identical to the exact results (exact results are given by enumerating all the configurations of the FVS, which is fortunately possible in this case because the system size is small). 
The figure also indicates that the BP and DC results are much worse than those in random graphs. The reason is that although the average degree of the Karate club $c=2.29$ is small, the network contains many short loops, which make Bethe approximation less accurate than in random graphs even with DC corrections. This effect turns out to be more severe for BP and DC on the political blogs networks as shown in Fig.~\ref{fig:correlation}(d), due to the power-law degree distribution and relatively large average degree $c=11.21$. But we can see the FVS method is less affected by the large degree and gives significantly more accurate correlation estimates than the other two methods. 

\subsection{Spin glasses on $2$D lattices}
In this section we consider statistical mechanics on the graphs with structured sparsity. A classical example is the $2$D ferromagnetic Ising model. Without external fields, the $2$D Ising model is on a planar graph, so we can compute an exact solution of the free energy using e.g. the Kac-Ward formula~\cite{kac1952combinatorial} and use it for evaluating our algorithm. On this problem we set a baseline using variational autoregressive network using convolution neural networks~\cite{wu2019solving}. Convolution networks are originally proposed to extract relevant features from $2$D data such as images~\cite{DLbook, pixelcnn}, and have been shown to give much better results than VAN without using convolution networks on the $2$D Ising model~\cite{wu2019solving}.

\begin{figure}[t]
\centering
\includegraphics[width=0.6\columnwidth]{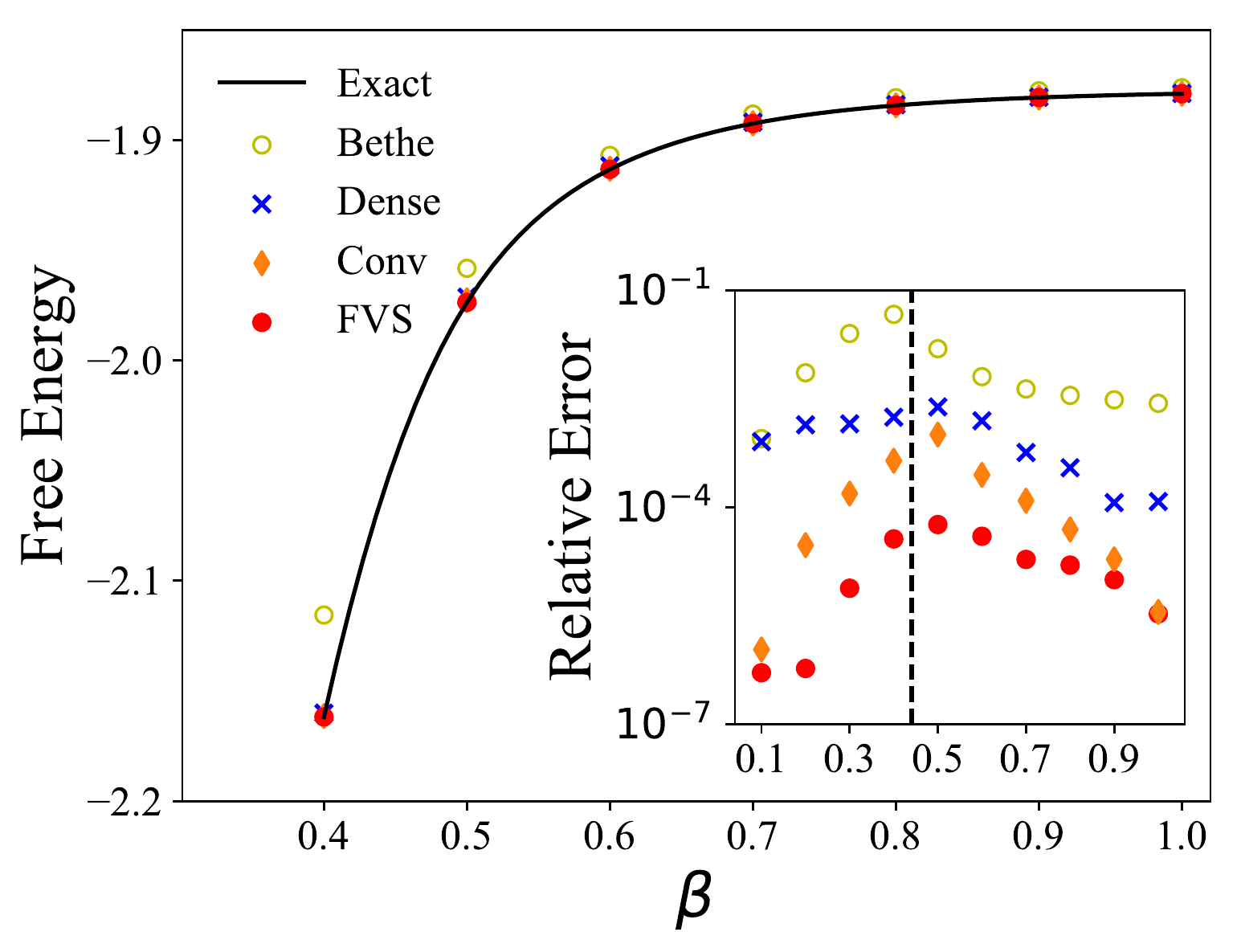}
\caption{\label{fig:lattice} Free energy (per spin) of the Ising model on $16\times 16$ lattice with open boundary condition obtained by our method (FVS), original two versions proposed in~\cite{wu2019solving} (Dense, Conv) and belief propagation (Bethe) with their relative errors to the exact solution~\cite{kac1952combinatorial}.  The vertical dashed line in the inset represents the phase transition point of an infinite system ($\beta = 0.4406868$).}
\end{figure}

In Fig.~\ref{fig:lattice} we plot the variational free energy given by our method at different temperatures, compared with other methods. The relative error to the exact solution is plotted in the inset of this figure. We can see clearly that in all temperature regime, our FVS method produces errors that are several orders of magnitude smaller than all the other methods, including the VAN based on convolution neural networks,  particularly on the paramagnetic-ferromagnetic transition point with $\beta\approx 0.4406868$ where the system has long-range correlations. 
Moreover our FVS based method employs much less parameters ($22532$) than the convolution networks ($714113$, details about model parameters are listed in appendix), indicating that our method is more effective than convolutions in characterizing the internal structure of the sparse statistical physics systems.

\begin{figure}[t]
\centering
\includegraphics[width=0.8\columnwidth]{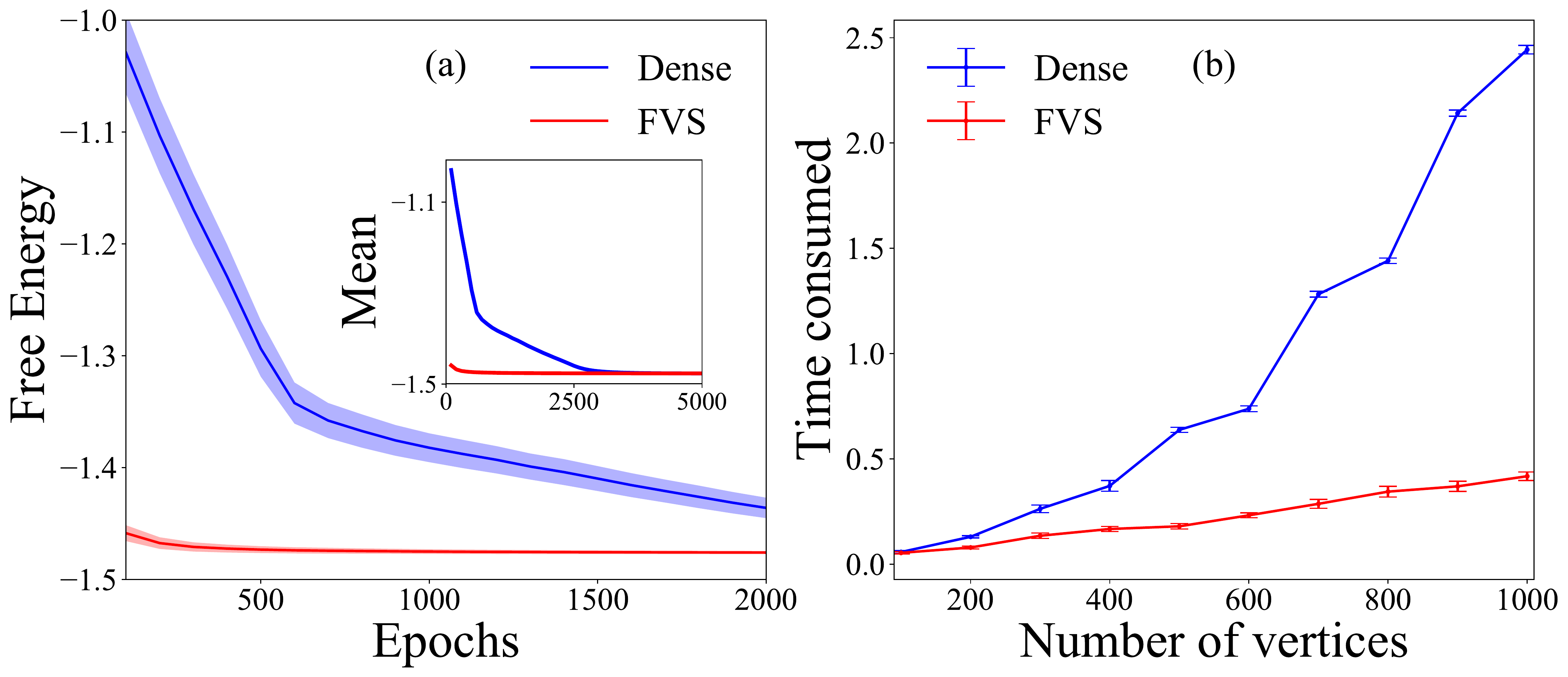}
\caption{\label{fig:accuracy}  (a) Evolution of the variational free energy through training. Solid lines represent mean of $E(\mathbf s)+\frac{1}{\beta}\ln{q_{\theta}(\mathbf s)}$, and shaded areas represent standard derivation. In the inset, a longer time scale of $5000$ epochs is illustrated; (b) Time used for one epoch (training step) in seconds. Each point is the average value over $100$ instances. Here FVS is our approach while Dense represent densely connected VAN in~\cite{wu2019solving}.}
\end{figure}

In Fig.~\ref{fig:accuracy} we compare as well the learning process as well as running time (in seconds) of our method and the VAN running on the whole graph~\cite{wu2019solving}. The figure illustrates that our method converges faster, and each learning step (epoch) takes significant less time in computation, hence is much more efficient than VAN operating on the whole graph.

\subsection{Inference problems on sparse graphs}

Apart from statistical mechanics problems, our method can also be adopted on statistical inference problems where the Bayesian inference formula corresponds to the Boltzmann distribution~\cite{iba1999nishimori,zdeborova2016statistical}. In this work we pick up the inference problem in the censored block model (CBM)~\cite{saade2015spectral} as an example. The CBM is a variant of the famous Stochastic Block Model~\cite{wang1987stochastic} in the field of community detection in networks. It asks to recover the hidden (or planted) group assignment (i.e. a configuration of binary variables) from censored edge measurements on a sparse graph. Given a graph $\mathcal {G}(\mathcal {V}, \mathcal {E})$ with $n$ nodes carrying binary variables $\sigma_i \in \{-1, +1\}$, each number represents a group, we have a planted partition $\hat{\sigma}$ which represents the ground truth of community. Our task here is to recover the planted configuration from the information of censored edges, which are depicted by edge labels $J_{ij}=\pm 1,(i,j)\in \mathcal E$ drawn from a distribution:
%
\begin{equation}
    P(J_{ij}|\hat{\sigma}_i, \hat{\sigma}_j) = (1-\alpha) \delta(J_{ij} - \hat{\sigma}_i \hat{\sigma}_j) + \alpha \delta(J_{ij} +\hat{\sigma}_i \hat{\sigma}_j) \; ,
\end{equation}
where $\alpha$ is the noise parameter,  ranging from $0$ (noiseless) to $0.5$ (informationless); $\delta(x)$ is the Kronecker symbol such that $\delta(x)=1$ if $x=0$ and $\delta(x)=0$ if $x \neq 0$.  The corresponding a posterior distribution is
\begin{equation}
    \label{eq:posterior}
    P(\bm{\sigma}|\bm{J}) = \frac{1}{Z}e^{\beta_0\sum\limits_{(ij)\in E} J_{ij}\sigma_i\sigma_j} \; ,
\end{equation}
which is exactly the form of Ising spin glass with $\beta_0 = 0.5\log{(1-\alpha)/\alpha}$ acting as inverse temperature. In this case, we use our FVSVAN method to estimate the posterior distribution and then infer the group assignment using the posterior distribution. The performance of algorithms can be evaluated by the fraction of correctly inferred group labels, the so called \textit{Fraction overlap}
\begin{equation}\label{eq:fraction_overlap}
    F(\bm{\hat{\sigma}}, \bm{\sigma}) = \frac{1}{n}\max\biggl(\sum\limits_i \delta(\hat{\sigma}_i - s_i), \sum\limits_i \delta(\hat{\sigma}_i + s_i)\biggr) \; ,
\end{equation}
where $s_i$ represents the inferred assignment of nodes $i$.
It has been shown in~\cite{saade2015spectral} that on large random graphs, the belief propagation algorithm as well as the non-backtracking~\cite{krzakala2013spectral} and Bethe-Hessian based algorithm are optimal in the sense that works all the way down to the theoretical detectability transition, analogous to the stochastic block model~\cite{decelle2011asymptotic}. 

\begin{figure}[t]
\centering
\includegraphics[width=0.7\columnwidth]{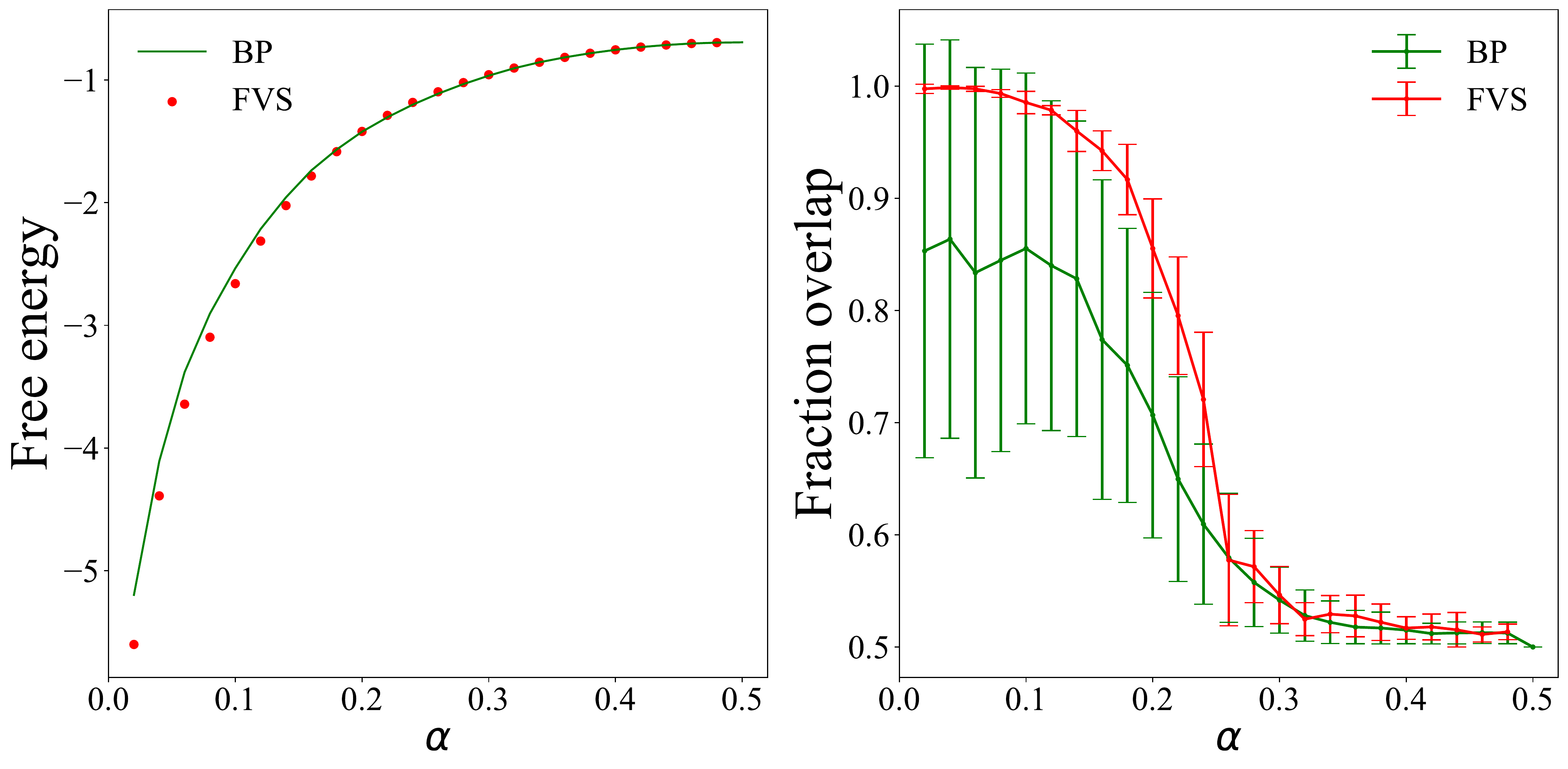}
\caption{\label{fig:sw}Inference in censored block model on a small-world network with $n=1000$ and $p=0.2$ versus the noise parameter $\alpha$ of censoring. (a) Free energy (per node, the smaller the better) of belief propagation (BP) and our method (FVS). (b) Fraction overlap~\eqref{eq:fraction_overlap} comparison, the FVS points are averaged over $10$ instances while the BP results are averaged over $300$ instances due to the large error bars.}
\end{figure}

We conduct experiments using the censored block model defined on small world networks (the Watts-Strogatz model~\cite{watts1998collective}), the results are shown in Fig.~\ref{fig:sw}. We can see from the left panel of the figure that the variational free energy computed by the FVSVAN is always smaller than that of BP. The right panel indicates that the accuracy of inference (Fraction overlap~\eqref{eq:fraction_overlap}) given by our method is much higher than BP. 
In the regime where the noise parameter $\alpha$ is large, the system is in the non-detectable phase~\cite{saade2015spectral,decelle2011asymptotic}and the censored edges carry almost no information about the planted partition. In this case we can see that both BP and our method reports overlap of $0.5$, basically a random guess on the group assignment.
While when the noise parameter $\alpha$ is low, where the inference of the planted partition is supposed to be easy because edge measurements $J_{ij}$ carry enough information about the ground-true planted partition. However what we can observer from Fig.~\ref{fig:sw} is that, the overlap obtained by BP is quite low even when the noise level is low, while our method can successfully recover the planted configuration, reporting high overlap. The reason for this phenomenon is that although the graph is sparse, BP still suffers from diverging issue due to small loops in the small-world networks, thus is not able to give an optimal estimate of marginals about group assignments and produces large error bars. Clearly our method can overcome the issue of BP and give an accuracy reconstruction when the edges measurements contain information about the planted partition.

%

\section{Conclusion and discussions\label{sec:con}}

We have introduced a new approach for solving statistical mechanics on sparse graphs. Our approach treats the random sparsity (such as random graphs and real-world networks) and the structured sparsity (such as lattices) in the same way, by extracting a small feedback vertex set, mapping the original problem to a statistical mechanics problem on the FVS with an effective energy, then solving it using neural network based variational method. We have illustrated using extensive numerical experiments that our method significantly outperforms existing methods on both spin glass problems and statistical inference problems on various of sparse graphs.

In this work we have adopted the feedback set to compress the sparse structures in graphs. We note that recently there is an interesting approach, graph convolution neural network~\cite{kipf2016semi}, which adapts convolution neural networks from lattices to sparse graphs, for some machine learning tasks such as the semi-supervised classifications. We have tested that naively applying the graph convolution (basically the message passing rule using neighborhood information) in the VAN works much worse than our FVSVAN approach. It would be interesting to study how to effectively combine the graph convolution approaches and the variational approach for solving statistical mechanics.

A potential application of our approach will be combinatorial optimization problems and constraint satisfaction problems defined on sparse graphs, such as vertex cover~\cite{Zhang2009a} and satisfiability problems~\cite{Mezard2002}, which can be converted to a statistical mechanics problem with inverse temperature $\beta\to\infty$.
We will put this into future work.

\begin{acknowledgements}
We want to thank Dian Wu and Lei Wang for helpful discussions, Alfredo Braunstein and Giovanni Catania for kindly running DC simulations for us.
H.J.Z. is supported by the National Natural Science Foundation of China (grant numbers 11975295 and 11947302) and the Chinese Academy of Sciences (grant number QYZDJ-SSW-SYS018).
P.Z. is supported by Key Research Program of Frontier Sciences, CAS, Grant No. QYZDB-SSW-SYS032, and Project 11947302 and 11975294 of National Natural Science Foundation of China. 
\end{acknowledgements}

\nocite{*}

\bibliography{apssamp}



\appendix

\newpage
\section{\label{app:VAN} Variational autoregressive networks}

The variational mean field methods  approximate the Boltzmann distribution using some tractable variational distributions from certain families. Due to the limitations of computing the variational free energy, the variational distributions are usually not very representative. The recent developments of deep neural networks are very helpful to offer much more representative variational distributions. The universal approximation theorem~\cite{DLbook} ensures that a neural network containing a large-enough hidden layer can approximate any continuous function. This means that a deep neural network might be used to approximate the Boltzmann distribution very accurately. This idea has recently been used in VAN for building a neural-network-based variational mean-field method for statistical mechanics problems~\cite{wu2019solving}, which factorizes the variational distribution by writing the joint probability as the product of conditional probabilities
\begin{equation}
    q(\bm{s}) = \prod\limits_{i=1}^{N}q(s_i|\bm{s}_{<i}) \; ,
    \label{eq:factorization}
\end{equation}
where $\bm{s}_{<i} \equiv (s_1, \ldots, s_{i-1})$ represents the state of all variables with indices less than $i$. Here we see that, for the variable with index $i$, its state $s_i$ depends on the variables with indices $j < i$ but not on those of the variables with indices $k > i$. This is called the autoregressive property in the machine learning community.
From the viewpoint of graphical models, Eq.~(\ref{eq:factorization}) connects all variables through their indices to a complete directed graph (suppose each variable is represented by a vertex). This connection pattern is universal for all graphical models but it makes no conditional independence assumptions about the variables.

Conditional probabilities $q(s_i|\bm{s}_{<i})$ then will be parameterized by carefully designing neural networks to represent the many-body interactions of the variables. Without loss of generality, for the simplest case of one layer autoregressive network without bias, the outputs can be write as
$$
s'_i = \sigma\Bigl( \sum_{j<i}W_{ij}s_j\Bigr) \; ,
$$
where $\sum_{j<i}W_{ij}$ is achieved by adding mask at weights $W_{ij}$. Since the sigmoid function $\sigma(\cdot)$ limits the output to be in the range of $(0,1)$, the output $s'_i$ can be naturally interpreted as a probability. In this case, the conditional probability can be written as
\beq
 q(s_i|\bm{s}_{<i}) = (s'_i)^{s_i} (1-s'_i)^{(1-s_i)} \; ,
   \label{eq:conditional}
\eeq
which is a Bernoulli distribution, with $s'_i = p(s_i=+1|\bm{s}_{<i})$. And the variational joint distribution becomes 
\beq
    q_{\theta}(\bm{s}) = \prod_{i=1}^{N}q_{\theta}(s_i|\bm{s}_{<i}) \; ,
    \label{eq:varjoint}
\eeq
where $\theta$ denote the set of all the network parameters like weights and bias. 

In variational methods, the parameters $\theta$ are adjusted to make the variational probability distribution $q_{\theta}(\bm{s})$ as close to the equilibrium  Boltzmann distribution as possible. The distance between two probability distributions is usually quantified by the  Kullback-Leibler (KL) divergence as~\cite{mackay2003information}
\beq
D_{\mathrm{KL}}\bigl(q \| p\bigr)=\sum_{\bm{s}} q(\bm{s}) \ln \Bigl(\frac{q(\bm{s})}{p(\bm{s})}\Bigr)
=\beta\bigl( F_{q}-F \bigr) \; ,
\label{eq:KL}
\eeq
where
\beq
    F_q = \sumup_{\bm{s}} q_{\theta}(\bm{s}) \Bigl[ E(\bm{s})+ \frac{1}{\beta}\ln q_{\theta}(\bm{s}) \Bigr]
\eeq
is the variational free energy, and $F$ is the true free energy. Since the KL divergence is positive, minimizing the KL divergence is equivalent to minimizing $F_q$ to its lower bound $F$. Thus the  variational free energy is set to be the loss function of the VAN. Here $E(\bm{s})$ is the energy function of the given model, and $\ln q_{\theta}(\bm{s})$ can be calculated combining Eq.~(\ref{eq:conditional}) and Eq.~(\ref{eq:varjoint})
\beq
    \ln q_{\theta}(\bm{s}) = \sumup_{i=1}^N \bigl[ s_i s'_i + (1-s_i)(1-s'_i) \bigr] \; .
   \label{eq:entropy}
\eeq
Minimizing $F_q$ requires optimization over the network parameters $\theta$. In our work the score function estimator~\cite{williams1992simple} is adopted to calculate the gradient of the variational free energy with respect to the network parameters:
\begin{equation}
\nabla_{\theta} F_q = 
\sum\limits_{\bm{s}} q_\theta(\bm{s}) 
\biggl\{ \Bigl[ E(\bm{s}) + \frac{1}{\beta} \ln q_{\theta}(\bm{s})\Bigr]  \nabla_{\theta} \ln q_{\theta}(\bm{s}) \biggr\} \; .
 \label{eq:gradients}
\end{equation}

Then, the parameters can be updated using a gradient based optimization such as the stocahstic gradient descent, or Adam ~\cite{kingma2014adam}, in order to make the variational free energy as smaller as possible. After training, the variational distribution $q_{\theta}(\bm{s})$ will be a good approximation to the Boltzmann distribution $p(\bm{s})$.

\begin{figure}[t]
\centering
\includegraphics[width=0.7\columnwidth]{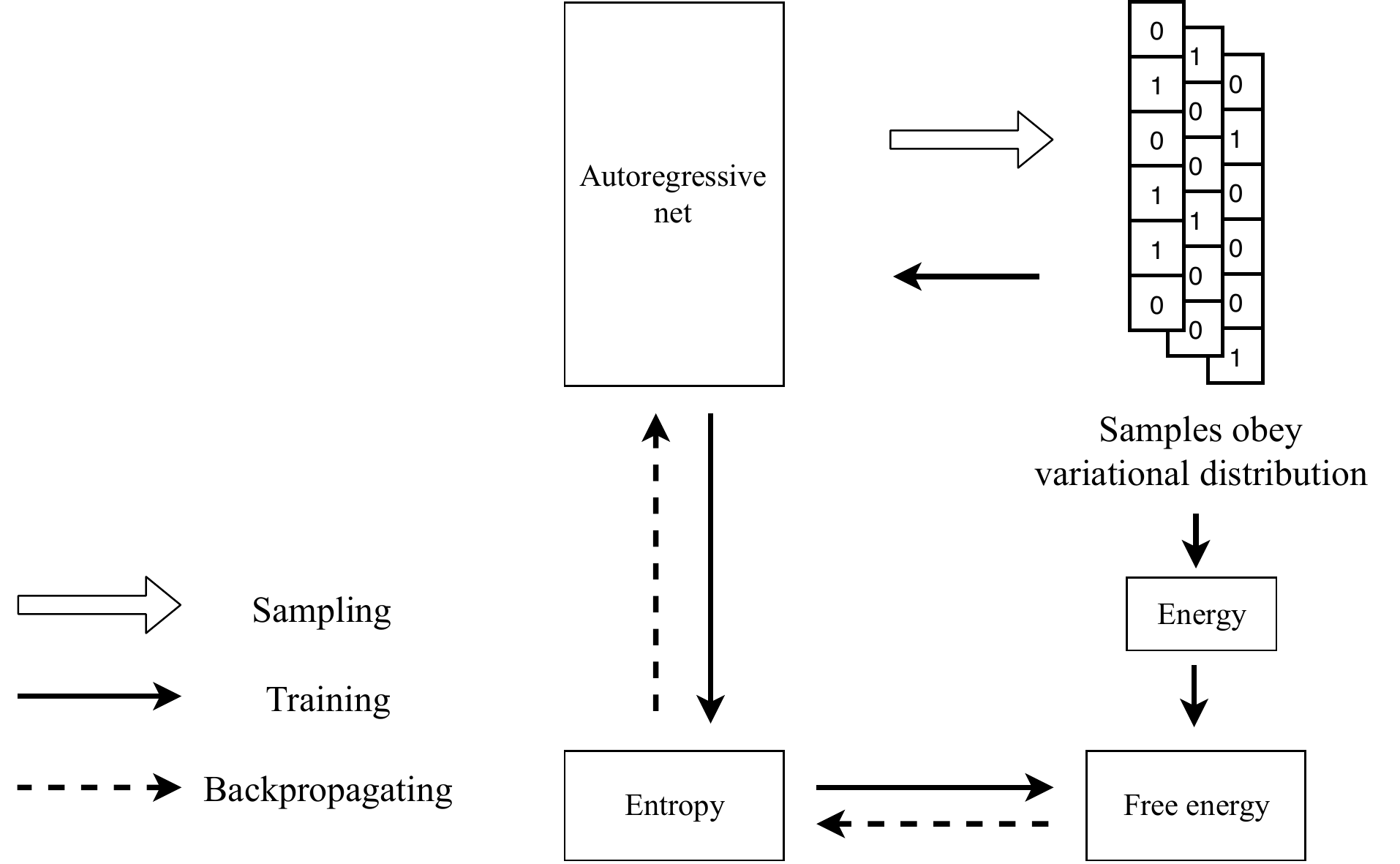}
\caption{Workflow of a variational autoregressive network}
\label{fig:workflow}
\end{figure}

The work flow of a VAN is shown in Fig.~\ref{fig:workflow}. At first, an autoregressive network needs to be constructed according to factorization of the joint probability. Then we need to get samples from the autoregressive network following the variational distribution through vertex by vertex sampling in a given ordering, the probability of next vertex given former vertices is given by Eq.~(\ref{eq:conditional}).
Then we can calculate energy $E(\bm{s})$, entropy $\ln q_{\theta}(\bm{s})$ and variational free energy $F_q$ using the samples, and estimate the gradients of $F_q$ using back propagation and update network parameters $\theta$ according to Eq.~(\ref{eq:gradients}).

\section{\label{app:deduction} Example of free energy computation on Ising model}

We consider the Ising spin glass as a representative model, in which the energy function is defined as 
\begin{equation}    E(\bm{\sigma})=-\sum_{(ij)\in\mathcal {E}}J_{ij}\sigma_i\sigma_j-\sum_i \sigma_i\theta_i \; ,
\end{equation}
where $\bm{\sigma}$ is a spin state of the system,  $\mathcal{E}$ denotes the set of edges, $J_{ij}$ is the coupling constant between two vertices $i$ and $j$ of the Ising model, and $\theta_i$ is the external field acting on vertex $i$. 

Given a configuration $\bm{s}$ of the FVS, the resulted effective external fields acting on the vertices out of the FVS are computed as
\begin{equation}
 \label{eq:hi0For}  
 h^0_i = \sum_{j\in \Omega\bigcap n(i)}\beta J_{ij} s_j \; ,
\end{equation}
where $n(i)$ is the set of neighbors of vertex $i$. Then the energy of the forest $\mathcal{T}$ complementary to the FVS will be
\begin{equation}
   E(\bm{t}) = -\sum_{(ij)\in\mathcal {E}_{\mathcal{T}}}J_{ij}t_i t_j-\sum_{i\in\mathcal{T}} t_i h_i \; .
\end{equation}
Here $\mathcal {E}_{\mathcal{T}}$ denotes the set of edges of the  forest $\mathcal{T}$, and $h_i = \theta_i + h_i^0$ is the overall external field on vertex $i$ of the forest.

When performing leaf removal (which removes all leaves of this graph recurrently), the whole forest will be hierarchized to levels with leaves on top and roots on bottom. Our leaf removal ordering is constructed by appending these levels from top to bottom to ensure that after removing vertices in upper level, vertices in the next level will all be leaves.

When a leaf $i$ is removed from the forest, the partition function of $\mathcal{T}$ can be written as
\beq
\begin{aligned}
    &Z(\bm{s}) = \sumup_{\bm{t}\in\mathbf{T}} e^{-\beta E(\bm{s}, \bm{t})} = \sumup_{t_i} e^{\beta t_i(\h_i + J_{ij}t_j)} \sumup_{\bm{t}_{\backslash i}}e^{-\beta E_{\backslash i}(\bm{t}_{\backslash i})} \\
    & = 2\cosh{\beta(J_{ij}t_j+\h_i)} \sumup_{\bm{t}_{\backslash i}}e^{-\beta E_{\backslash i}(\bm{t}_{\backslash i})} \\
    & = 2 \sqrt{\cosh{\beta(J_{ij}+\h_i)}\cosh{\beta(-J_{ij}+\h_i)}} \\
    & \exp{\bigg[ \frac{t_j}{2} \ln \big(\frac{\cosh{\beta(J_{ij}+h_i)}}{\cosh{\beta(-J_{ij}+h_i)}}\big) \bigg]}\sumup_{\bm{t}_{\backslash i}}e^{-\beta E_{\backslash i}(\bm{t}_{\backslash i})} \\
    & = 2 \sqrt{\cosh{\beta(J_{ij}+\h_i)}\cosh{\beta(-J_{ij}+\h_i)}} \sumup_{\bm{t}_{\backslash i}}e^{\beta h'_jt_j}e^{-\beta E_{\backslash ij}(\bm{t}_{\backslash i})} \\
    & = 2 \sqrt{\cosh{\beta(J_{ij}+\h_i)}\cosh{\beta(-J_{ij}+\h_i)}} Z_{\backslash i} \; ,
\end{aligned}
\eeq
where $E_{\backslash i}(\bm{t}_{\backslash i})$ represents energy with vertex $i$ and its interaction excluded,  $E_{\backslash ij}(\bm{t}_{\backslash i}) = E_{\backslash i}(\bm{t}_{\backslash i}) - h_jt_j$ and $h'_j$ is the external field that combines the original external field of vertex $j$ and a external field vertex $i$ exert on $j$.

Thus when a leaf vertex $i$ is removed from the graph, it gives a factor $2 \sqrt{\cosh{\beta(J_{ij}+\h_i)}\cosh{\beta(-J_{ij}+\h_i)}}$ to the partition function and change the external field of its neighbor $j$ is expressed as
\beq
\h'_j = \h_j + \frac{1}{2\beta}\ln\bigg(\frac{\cosh{\beta(J_{ij}+\h_i)}}{\cosh{\beta(-J_{ij}+\h_i)}} \bigg) \; .
\label{eq:field}
\eeq
For the root vertex $k$, the factor is $2\cosh{\beta h_k}$ since it has no neighbors when removed. 

Multiplying these factors together and combining Eq.~\eqref{eq:sumup} and Eq.~\eqref{eq:tree energy}, the effective energy of the FVS can be analytically expressed as


 \begin{align}\label{eq:effective_energy}
\widetilde E(\mathbf s)&=E_\Omega(\mathbf s)+\frac{1}{\beta}\left[  \sum_{(i,j)\in \mathcal{E}_{\mathcal{T}}}\ln\sqrt{4\cosh\beta(J_{ij}+\h_i)}
    +\sum_{(i,j)\in \mathcal{E}_{\mathcal{T}}}\ln\sqrt{4\cosh\beta(-J_{ij}+\h_i)}+\sum_{i\in \mathcal{R}}\ln(2\cosh(\beta \h_i))\right] \; .
 \end{align}
 
Here $\mathcal{R}$ denotes set of roots of the remaining forest, and $\h_i$ is the effective field acting on vertex $i$ when $j$ is a neighbor of $i$:

\section{\label{app:experiment} Experimental settings}

In the first experiment, we compare correlations of three methods based on spin glass models. 
Since connected correlations calculation of MCMC is straightforward, here we only introduce how to calculate connected correlations of other two methods.

For the belief propagation, we adopt the scheme of~\cite{zhou2015spin} by propagating cavity fields $h_{i\rightarrow j}$. As long as cavity fields reaching their fixed point, we can calculate physical quantities based on them. The magnetism of vertex $i$ is
\beq
    m_i = \sumup_{j\in \partial i} \arctanh[\tanh(\beta J_{ij})\tanh(\beta h_{j\rightarrow i})] \; ,
\eeq
where $\partial i$ represents all neighbors of vertex $i$. And connected correlation of $i$ and $j$ is
\beq
    \la s_i s_j \ra_c = 
    \frac{e^{\beta J_{ij}}\cosh{\beta h^+_{ij}} - e^{-\beta J_{ij}}\cosh{\beta h^-_{ij}}}{e^{\beta J_{ij}}\cosh{\beta h^+_{ij}} + e^{-\beta J_{ij}}\cosh{\beta h^-_{ij}}}
    -m_im_j \; ,
\eeq
where $h^+_{ij} = h_{i\rightarrow j} + h_{j\rightarrow i}$ and $h^-_{ij} = h_{i\rightarrow j} - h_{j\rightarrow i}$.

For our method, calculations can be a little obscure since we only have samples of FVS vertices but correlations of all edges are needed. But we can start from the definition of correlations
\begin{align}
   \la s_i s_j \ra &= \sumup_{s_i s_j}s_i s_j p(s_i s_j) \nonumber\\
    &=\sumup_{s_i s_j}s_i s_j\frac{\sum_{\bm{s}\backslash s_is_j}e^{-\beta E(\bm{s})}}{Z} \nonumber\\
    & = \sumup_{s_i s_j} s_i s_j \frac{ \sum_{\bm{s}_{\text{fvs}}} e^{-\beta \Tilde{E}(\bm{s}_{\text{aug}})}}{\sum_{\bm{s}_{\text{fvs}}} e^{-\beta \Tilde{E}(\bm{s}_{\text{fvs}})}} \nonumber\\
    & = \frac{Z_{++}+Z_{--}-Z_{+-}-Z_{-+}}{\sum_{\bm{s}_{\text{fvs}}} e^{-\beta \Tilde{E}(\bm{s}_{\text{fvs}})}} \; ,
\end{align}
where $\bm{s}_{\text{aug}} = \bm{s}_\text{fvs+ij}$ is the state of FVS vertices plus vertex $i$ and $j$, $Z_{++} = \sum_{\bm{s}_{\text{fvs}}} e^{-\beta \Tilde{E}(\bm{s}_{\text{fvs}}|s_i=+1,s_j=+1)}$. Then both the numerator and denominator can be calculated using our method. $\la s_i\ra$ can be calculated similarly as
\begin{equation}\label{eq:marginal}
    \langle s_i \rangle = \frac{Z_{+}-Z_{-}}{\sum_{\bm{s}_{\text{fvs}}} e^{-\beta \Tilde{E}(\bm{s}_{\text{fvs}})}} \; ,
\end{equation}
where $Z_{+} = \sum_{\bm{s}_{\text{fvs}}} e^{-\beta \Tilde{E}(\bm{s}_{\text{fvs}}|s_i=+1)}$.
Connected correlations of entire graph can be calculated with expressions above.

\begin{table}[h]
\caption{\label{tab:hyper}
Hyperparamters of Fig.~\ref{fig:lattice}}
\begin{ruledtabular}
\begin{tabular}{cccc}
    & Dense & Convolution & FVS \\
    \hline
    Batch & $1\times 10^3$ & $1\times 10^3$ 
    & $1\times 10^4$ \\

    Net depth & 3 & 3 & 2  \\

    Net width & 4 & 64\footnote{For convolution, this number represent channels.} & 3 \\

    Max step & $1\times 10^4$ & $1\times 10^4$
          & $1\times 10^4$\\
          
    Learning rate & $1\times 10^{-3}$ & $1\times 10^{-3}$ &
    $1\times 10^{-3}$ \\
    Number of parameters\footnote{This quantity is not a hyperparameter, but we list it here for comparison.} & 1577216 & 714113 & 22532 \\
\end{tabular}
\end{ruledtabular}
\end{table}

In the second experiment, our method, two architectures from original version of VAN, and the belief propagation are implemented on a $16\times 16$ 2D ferromagnetic Ising model with open boundary conditions. For dense and convolution architectures, we borrow hyperparameter settings from~\cite{wu2019solving}. All hyperparameters used in this experiment are listed in Table~\ref{tab:hyper}. Since all these methods have been fine-tuned for their best performance, there are some differences on the  hypereparameter settings. 

In the third experiment in the main text, two methods are implemented on random regular graphs to compare their running times and convergence speed. Hyperparamters like batch size, depth of the neural network, and learning rate used in Fig.~\ref{fig:accuracy}(a) are all the same for two methods, while the numbers of trainable parameters differs a lot: $63503$ for the FVS and $1002000$ for the dense VAN,  for a random graph of size $N=1000$. We see our method massively decreases the number of hyperparameters. The hyperparameters of Fig.~\ref{fig:accuracy}(b) are similar to Fig.~\ref{fig:accuracy}(a), with the only difference on the number of vertices.

In the fourth experiment, we run BP and our method on the censored block model defined on small world networks. The belief propagation is written as
\begin{equation}\label{eq:bp1}
  \psi_{\sigma_i}^{i\rightarrow j} =\frac{\prod_{k\in \partial i\setminus j} \sum_{\sigma_k}e^{\beta_0 J_{ik}\sigma_i\sigma_k}\psi_{\sigma_k}^{k\rightarrow i}}{Z^{i\rightarrow j}} \; ,
\end{equation}
where $\psi_{\sigma_i}^{i\rightarrow j}$ is the probability of node $i$ being as spin or group $\sigma_i$ with node j removed from the graph. After Eq.~\eqref{eq:bp1} converged or iterated 2000 iterations, we evaluate marginal of every variable as
\begin{equation}\label{eq:bp2}
  \psi_{\sigma_i}^{i} =\frac{\prod_{k \in \partial i} \sum_{\sigma_k}e^{\beta_0 J_{ik}\sigma_i\sigma_k}\psi_{\sigma_k}^{k\rightarrow i}}{Z^{i}} \; .
\end{equation}
For our method, one-layer VAN will be enough to approximate the posterior distribution Eq.~\eqref{eq:posterior}, which contains $l(l+1)/2$ parameters ($l$ is the size of FVS). After training, samples of FVS will be obtained but they can not used to directly infer ground truth. This is because unlike other methods, there is no spontaneous symmetry breaking of VAN, which means samples of VAN can be divided into two equally probable part with opposite orientation so all marginals will tend to be $0$. Thus manually symmetry breaking need to be performed to these samples to break the $\mathbb{Z}_2$ symmetry of censored block model. We pick a sample $\bm{\sigma_l}$ with lowest energy as a target, and flip all other samples according to inner products:
\begin{equation}
    \bm{\widetilde{\sigma_j}} = \text{sgn}(\bm{\sigma_j} \cdot \bm{\sigma_l}) * \bm{\sigma_j} \; .
\end{equation}
After flipping, marginals can be calculated. For FVS vertices, marginals can be calculated from VAN samples, while for non-FVS vertices, Eq.\eqref{eq:marginal} can be used. Then inferred assignment of node $i$ will be
\begin{equation}
    s_i = \text{sgn}(\langle \widetilde{\sigma_i} \rangle) \; ,
\end{equation}
then the fraction overlap is calculation according to Eq.~\eqref{eq:fraction_overlap}.

\end{document}